\documentstyle[aps,epsf]{revtex}

\def\be{\begin{equation}}
\def\ee{\end{equation}}
\def\bea{\begin{eqnarray}}
\def\eea{\end{eqnarray}}

\def\t12h{\frac{\theta_{12}}{2}}

\def\eps{\varepsilon}

\def\r#1{(\ref{#1})}
\def\nn{\nonumber\\}

\def\ezp{{\eps^0}^\prime(q)}
\def\ezpp{{\eps^0}^{\prime\prime}(q)}
\def\ezppp{{\eps^0}^{\prime\prime\prime}(q)}

\def\dr{\Delta\rho}

\def\ezpp{{\eps^0}^{\prime\prime}(q)}

\def\NPB#1#2#3{{ Nucl. Phys.} {\bf B#1} (#2) #3}
\def\JPA#1#2#3{{ J. Phys.} {\bf A#1} (#2) #3}

\def\PRB#1#2#3{{ Phys. Rev.} {\bf B#1} (#2) #3}
\def\PRL#1#2#3{{ Phys. Rev. Lett.} {\bf #1} (#2) #3}

\def\CMP#1#2#3{{Comm. Math. Phys.} {\bf #1} (#2) #3}

\begin {document}

\preprint{}

\title{\bf Temperature Corrections to Conformal Field Theory}

\author{Fabian H.L. E\ss ler$^{(a)}$, Vladimir E. Korepin$^{(b)}$
and Franck T. Latr{\'e}moli{\`e}re$^{(a)}$}
\address{$^{(a)}$ Department of Physics, Theoretical Physics,
        Oxford University\\ 1 Keble Road, Oxford OX1 3NP, United Kingdom}
\address{$^{(b)}$ Institute for Theoretical Physics, SUNY at Stony
Brook, Stony Brook, NY 11794-3840, USA}
\maketitle
\begin{abstract}
\par
We consider finite temperature dynamical correlation functions in the
interacting delta-function Bose gas. In the low-temperature limit the
asymptotic behaviour of correlation functions can be determined from
conformal field theory. In the present work we determine the
deviations from conformal behaviour at low temperatures.
\end{abstract}

\vskip .5cm
PACS numbers: 05.30.-d, 05.30.Jp, 65.90.+i

\section{Introduction}
The calculation of finite temperature correlation functions is a
long-standing problem in the theory of integrable models. 
Apart from the obvious conceptual importance of the problem there are
many direct applications of the results to experiments on quasi-1D
materials like ${\rm KCuF_3}$ \cite{kcuf3} or ${\rm CuBenz}$
\cite{cubenz}, which are described by integrable models. Finite
temperature dynamical correlation functions in the systems have been
measured by Neutron scattering and NMR and it is highly desirable to
develop a method to calculate them exactly.

Important progress was made during the eighties when it was realized
that in gapless models Conformal Field Theory can be used to obtain
the low-temperature asymptotics of dynamical correlation functions
(see {\sl e.g.} \cite{cft}). However, models with a spectral gap as
well as higher temperatures in gapless models remain outside the scope
of the conformal approach. 

In a remarkable further development it became possible to determine
the behaviour of {\sl static} correlators at finite temperatures
through an ingenious mapping of integrable d-dimensional quantum
theories to d+1-dimensional integrable classical statistical models
\cite{tba}. However, dynamical correlation functions cannot presently
be calculated by this approach.

For integrable models with free fermionic spectra powerful methods to
calculate finite temperature dynamical correlation functions have been
available for some time \cite{vladb,freef}. Very recently the method of
\cite{vladb} was successfully extended to cases corresponding to {\sl
interacting} fermions \cite{slav,ks}. In particular, in \cite{ks} a formula
describing the exponential decay of correlations in the delta-function
Bose gas at finite temperatures was presented.
Said formula is implicit in the sense that it is written in terms of
solutions of certain nonlinear integral equations. The purpose of the
present work is to analyze these integral equations by both analytical
and numerical methods and present explicit expressions for the
correlation lengths describing the decay of correlations.

The outline of the paper is as follows. In section 1 we review some
relevant facts on the delta-function Bose gas. In section 2 we study
the special case of impenetrable bosons, in which particularly simple
expressions for the correlation lengths are obtained at low
temperatures. In section 3 we consider the general case of interacting
bosons and we conclude in section 4.

\section{Review of the $\delta$-function Bose gas}

The $\delta$-function Bose gas is one of the paradigms of exactly
solvable strongly correlated many-body problems in one spatial
dimension. It describes $N$ bosons interacting {\sl via} a repulsive
$\delta$-function potential of strength $c>0$. The Hamiltonian is
\begin{equation}\label{Ham1}
{\cal H}_N=-\sum_{j=1}^{N}\frac{\partial^2}{\partial x_j^2}
+2c\sum_{N\ge j>k\ge1}\delta(x_j-x_k).
\end{equation}
At the special value $c=\infty$ the model describes noninteracting
hard-core bosons and essential simplifications occur in the exact
solution. We refer to this case as ``impenetrable bosons''.
The second-quantized form of the model is known as the Quantum
Nonlinear Schr\"odinger equation and the Hamiltonian is expressed in
terms of the canonical Bose field $\psi(x)$ as
\begin{equation}\label{Ham2}
{H}=\int_{-\infty}^{\infty} dx
\left({\partial_x}\psi^{\dagger}(x)
{\partial_x} \psi(x)+
c\psi^{\dagger}(x)\psi^{\dagger}(x)\psi(x)\psi(x)\right).
\end{equation}

The model is solvable by Bethe Ansatz \cite{LL} and in what follows we
recall some important ingredients of the exact solution.
\begin{itemize}
\item{\bf Ground State and Excitations}

In momentum space a Pauli principle holds \cite{IK} and consequently
the zero temperature ground state is given by a filled Fermi sea of
negative energy pseudoparticles. At $c=\infty$ this is simply the
usual free-fermion ground state. The physics of the model is most
conveniently described in terms of a rapidity variable $\lambda$,
which is related to the momentum $k$ by 

\begin{equation}\label{mom0}
k(\lambda)=\lambda+\int_{-q}^{q} \theta(\lambda-\mu)
\rho^0_t(\mu)\,d\mu\ ,
\end{equation}
where $\theta(\lambda)=i\ln\left(\frac{ic+\lambda}{ic-\lambda}\right)$.
%
The density $\rho^0_t(\lambda)$ of pseudoparticles in the ground state is
described by the integral equation
\begin{equation}\label{rho0}
\rho^0_t(\lambda)-\frac{1}{2\pi}\int_{-q}^{q} K(\lambda,\mu)
\rho^0_t(\mu)\,d\mu=\frac{1}{2\pi}.
\end{equation}
Here $q$ is the rapidity corresponding to the Fermi momentum $k_F$ and
\begin{equation}\label{kernel}
K(\lambda,\mu)=\frac{2c}{c^2+(\lambda-\mu)^2}.
\end{equation}

Excitations over the ground state can be either particles or
holes. The particle energy as a function of the rapidity is given by
\begin{equation}\label{eps0}
\eps^0(\lambda)-\frac{1}{2\pi}\int_{-q}^{q} K(\lambda,\mu)
\eps^0(\mu)\,d\mu=\lambda^2-h\ ,
\end{equation}
where $h>0$ is the chemical potential. We note that the excitation
energy vanishes on the Fermi surface $\eps^0(\pm q)=0$. 
The dependence of the integration boundary $q$ on the chemical
potential $h$ is determined from the condition 
\be
\eps^0(\pm q)= 0\ .
\ee
In the impenetrable case the excitation spectrum is given in terms of 
free-fermionic particle-hole excitations.
The Fermi velocity is defined as usual to be
\begin{equation}\label{velocity1}
v_F=\left.\frac{\partial\varepsilon(\lambda)}
{\partial k(\lambda)}
\right|_{\lambda=q}=\frac{{\eps^0}^\prime(q)}{k'(q)}
	=\frac{{\eps^0}^\prime(q)}{2\pi\rho^0_t(q)}
\end{equation}
where the prime denotes differentiation with respect to $\lambda$.
The derivative of momentum with respect to rapidity on the Fermi
surface is called dressed charge and is related to the density of
states on the Fermi surface
\begin{equation}\label{charge}
Z=2\pi\rho^0_t(q).
\end{equation}
\item{\bf Asymptotics of Correlation Functions}

The asymptotic behaviour of correlation functions at very low
temperatures can be determined from the exact finite-size spectrum
by conformal field theory techniques \cite{vladb,alex}. The
result is found to be 

\begin{equation}\label{asym2}
\langle\psi(0,0)\psi^\dagger(x,t)\rangle_T
\begin{array}{c}
{}\\
\longrightarrow\\
\mbox{\raisebox{2mm}{$x\to\infty\atop t\to\infty$}}
\end{array}
\exp\left\{-\frac{2\Delta^+\pi T}{v_F}
|x-v_Ft|-\frac{2\Delta^-\pi T}{v_F}|x+v_Ft|\right\}.  
\end{equation} 

The conformal dimensions $\Delta^\pm$ are related to the dressed
charge by
\begin{equation}\label{confdim}
2\Delta^+=2\Delta^-= \frac{1}{4Z^2}.
\end{equation}
In the framework of the Luttinger liquid approach these results were
obtained by F.~D.~M. Haldane in \cite{H}. 

In \cite{ks} the following formula describing the exponential decay of
correlations at {\sl any} temperature was derived from the determinant
approach to quantum correlation functions \cite{vladb}
\begin{equation}\label{asym3}
\langle\psi(0,0)\psi^\dagger(x,t)\rangle_T
\begin{array}{c}
{}\\
\longrightarrow\\
\mbox{\raisebox{2mm}{$x\to\infty\atop t\to\infty$}}
\end{array}
\exp\left\{\frac{1}{2\pi}\int_{-\infty}^{\infty}
\frac{d\lambda}{2\pi\rho_t(\lambda)}
|x-v(\lambda)t|\ln\left|\frac
{e^{\frac{\eps(\lambda)}{T}}-1}
{e^{\frac{\eps(\lambda)}{T}}+1}
\right|\right\}=: \exp\left(\chi\right)\ .
\end{equation}
The functions $\eps(\lambda)$, $\rho_t(\lambda)$ and $v(\lambda)$ are
the finite-temperature equivalents of the dressed energy,
pseudoparticle density and Fermi velocity defined above. They are
solutions of the following integral equations \cite{YY}
\bea
\eps(\lambda)&=&\lambda^2-h-
\frac{T}{2\pi}\int_{-\infty}^{\infty}d\mu\ K(\lambda,\mu)
\ln\left(1+e^{-\frac{\eps(\mu)}{T}}\right)\ ,\nn
\rho_t(\lambda)&=&\frac{1}{2\pi}+\frac{1}{2\pi}
\int_{-\infty}^{\infty}d\mu\ K(\lambda,\mu)\
\frac{1}{1+e^{\frac{\eps(\mu)}{T}}}\ \rho_t(\mu)\ .
\label{inteq}
\eea
The velocity $v(\lambda)$ is given by
\begin{equation}\label{velocity2}
v(\lambda)=\frac{1}{2\pi \rho_t(\lambda)}
\frac{\partial\eps(\lambda)}{\partial\lambda}\ .
\end{equation}
Note that for $T\rightarrow 0$ these equations reduce to \r{eps0},
\r{rho0} and \r{velocity1} respectively. 
\end{itemize}

\section{Impenetrable Bosons}
We first investigate the simpler case of impenetrable bosons. 
In the limit $c\rightarrow\infty$ \r{asym3} simplifies to \cite{vladb}
\begin{equation}\label{imp1}
\langle\psi(0,0)\psi^\dagger(x,t)\rangle_T
\begin{array}{c}
{}\\
\longrightarrow\\
\mbox{\raisebox{2mm}{$x\to\infty\atop t\to\infty$}}
\end{array}
\exp\left\{\frac{1}{2\pi}\int_{-\infty}^{\infty}
{d\lambda}\ |x-2\lambda t|\ \ln\left|\frac
{e^{\frac{\lambda^2-h}{T}}-1}
{e^{\frac{\lambda^2-h}{T}}+1}
\right|\right\} ,
\end{equation}
and we are left with performing a single integral. Expanding the
exponentially decaying part of the integrand as
\be
\ln\left|\frac
{e^{\frac{\lambda^2-h}{T}}-1}{e^{\frac{\lambda^2-h}{T}}+1}\right|
=-2\sum_{n=1}^\infty\frac{1}{2n-1}
\exp\left(-\left|\frac{\lambda^2-h}{T}\right|(2n-1)\right)
\ee
we obtain an expansion of \r{imp1} in terms of error functions and
exponential functions, which in turn yield an {\sl asymptotic}
low-temperature series on $\chi$ in powers of $T$.
The Fermi velocity is $v_F=2\sqrt{h}$, and we distinguish two cases.
\begin{itemize}
\item{\underbar{Space-like region ${x}/{t}>v_F$:}}

After some elementary calculations we find the following expansion 
for $\chi$ (as defined in \r{asym3})
\be
\chi\sim-\frac{4xT}{\pi v_F}\sum_{m=0}^\infty\frac{(4m)!}{(2m)!}
\left(\frac{T}{4h}\right)^{2m} (1-2^{-2-2m})\ \zeta(2m+2)\ ,
\ee
where $\zeta(x)$ is the Riemann zeta-function. Note that this
expansion breaks down as we approach $x/t\rightarrow v_F$.
The first few terms are
\be
\chi=-\frac{xT\pi}{2v_F}\left[1+\left(\frac{\pi T}{v_F^2}\right)^2
+14\left(\frac{\pi T}{v_F^2}\right)^4
+5049\left(\frac{\pi T}{v_F^2}\right)^6+{\cal O}(T^8)\right]\ .
\ee
The inverse correlation length thus has an asymptotic power series
expansion in {\sl odd} powers of temperature for $T\rightarrow
0$. We also see that in the space-like regime there is no exponential
decay with respect to the time $t$ at low temperatures as the
$t$-dependence only enters in terms such as $\exp({\rm const}/T)$
which are smaller than any power of the temperature $T$.

\item{\underbar{Time-like region ${x}/{t}<v_F$:}}

In the time-like regime we find that there are no power-law
corrections to the conformal result
\be
\chi=-\frac{\pi t T}{2} - \frac{2 t T}{\pi}
\exp\left(-[h-x^2/4t^2]/T\right) + \ldots
\ee
Thus, up to exponentially small corrections the conformal result is
exact in the time-like regime and there is exponential decay of
correlations with respect to $t$ only at small temperatures.
\end{itemize}

\begin{figure}[ht]
\begin{center}
\noindent
\epsfxsize=0.65\textwidth
\epsfbox{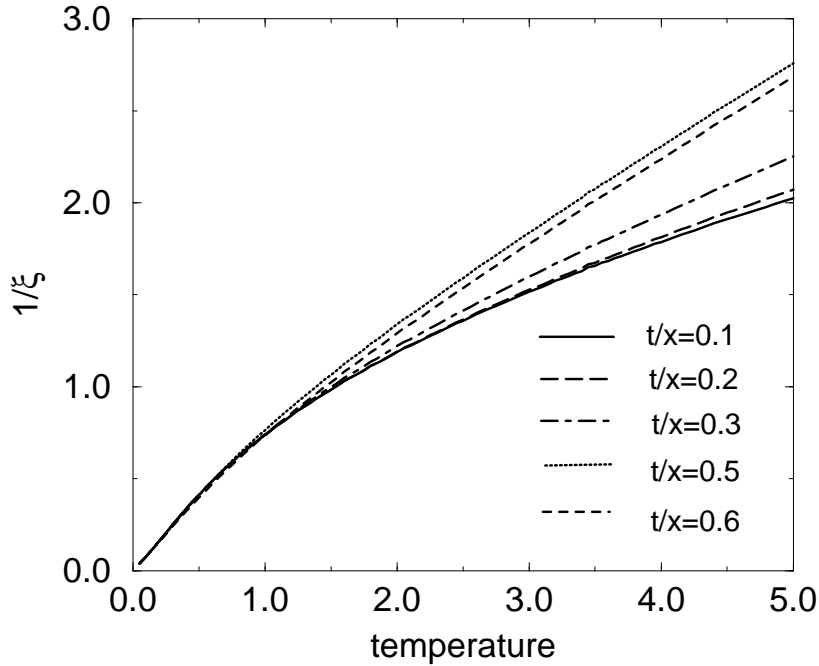}
\end{center}
\caption{\label{fig:c1000}%
Inverse correlation length as a function of temperature for
impenetrable bosons at $h=1$. 
} 
\end{figure}

For general finite temperatures one needs to resort to numerical
integration of \r{imp1}. If we define a correlation length
$\xi(T,h,\frac{t}{x})$ by 
\be
\chi=-\Theta(x-v_Ft)\frac{x}{\xi(T,h,\frac{t}{x})}
-\Theta(v_Ft-x)\frac{v_Ft}{\xi(T,h,\frac{t}{x})}\ ,
\label{length}
\ee
where $\Theta(x)$ is the Heaviside theta-function, we can study the
dependence of $\xi$ on $T$ and the ``direction'' ${t}/{x}$ by
determining $\chi$ numerically. From the above low-temperature
analysis we already know that at very low temperatures $\xi$ becomes
independent of ${t}/{x}$. In Fig~\ref{fig:c1000} we plot the
correlation length $\xi$ as a function of temperature for various
values of ${t}/{x}$ for the special values of chemical potential $h=1$. 
We clearly see the deviations from linear-T behaviour as temperature
increases. The dependence on $t/x$ becomes more pronounced at higher
temperatures. In the regions $tv_F\ll x$ and $x\ll v_Ft$ the
dependence on $t/x$ disappears as expected. We also see that the
exponential decay of correlations is fastest in the direction of the
``light-cone'' $x/t=v_F$. 

\section{Interacting Bosons}
We now consider the case of finite coupling constant $c<\infty$.
In order to derive an analytical
low-temperature expansion of the correlation length we first need to
perform an (asymptotic) expansion of $\eps(\lambda)$, $v(\lambda)$ and
$\rho_t(\lambda)$ in powers of $T$.

\subsection{Low temperature expansion of integral equations}

The expansion of $\eps(\lambda)$ in powers of $T$ is readily
established along the lines of \cite{taka}. We first note that the
function $\eps(\lambda)$ has precisely one zero for $\lambda >0$
\cite{vladb}. We denote the corresponding value of the spectral
parameter by $q_T$. As $\eps(\lambda)$ is a symmetric function this
then implies that $\eps(\pm q_T)=0$. From (\ref{inteq}) and
(\ref{eps0}) we find 
\bea
\eps(\lambda)-\eps^0(\lambda)&-&\frac{1}{2\pi}\int_{-q}^{q}d\mu\ 
K(\lambda,\mu)\ \left[\eps(\mu)-\eps^0(\mu)\right]\nn
&=&-\frac{T}{2\pi}\int_{-\infty}^\infty d\mu\ K(\lambda,\mu)\
\ln\left(1+\exp(-\bigg|\frac{\eps(\mu)}{T}\bigg|)\right)
-\frac{T}{2\pi}\int_{q}^{q_T}\!\!\!\!+\int_{-q_T}^{-q} d\mu\ K(\lambda,\mu)\ 
\ln\frac{1+\exp(-\frac{\eps(\mu)}{T})}
{1+\exp(-|\frac{\eps(\mu)}{T}|)}\ .
\label{diffe}
\eea
The last two terms in (\ref{diffe}) are easily evaluated by expanding
the integrand around $\pm q_T$. The leading contribution to the first term
in (\ref{diffe}) also comes from the regions $\mu=\pm q_T$ and we
again expand the integrand around these points and perform the
resulting integrals. This gives

\bea
\eps(\lambda)-\eps^0(\lambda)&-&\frac{1}{2\pi}\int_{-q}^{q}d\mu\ 
K(\lambda,\mu)\ \left[\eps(\mu)-\eps^0(\mu)\right]\nn
&=&-\frac{\pi}{12\eps^\prime(q)}\left[K(\lambda,q)+K(-\lambda,q)\right]T^2
+\frac{\eps^\prime(q_T)}{4\pi}\left[K(\lambda,q_T)+K(-\lambda,q_T)
\right](q_T-q)^2 +{\cal O}(T^3)\ .
\label{diffe2}
\eea
Setting $\lambda=q$ in (\ref{diffe}) we find that $q-q_T={\cal
O}(T^2)$ so that the last term in (\ref{diffe}) does not contribute to
${\cal O}(T^2)$. Putting everything together we then have

\bea
\eps(\lambda)&=&\eps^0(\lambda)-\frac{\pi}{12{\eps^0}^\prime(q)}T^2
u(\lambda)+{\cal O}(T^3)\ ,\nn
u(\lambda)&=&K(\lambda,q)+K(-\lambda,q)+\frac{1}{2\pi}
\int_{-q}^{q}d\mu\ K(\lambda,\mu)\ u(\mu)\ .
\label{eT}
\eea
It immediately follows that
\be
q_T=q+\frac{\pi}{12} \frac{u(q)}{[{\eps^0}^\prime(q)]^2} T^2 + {\cal
O}(T^3)\ .
\ee
The finite-temperature corrections to the pseudoparticle density 
$T^2\Delta\rho(\mu)=\rho_t(\mu)-\rho_t^0(\mu)$ 
can be determined in a similar way
\bea
\dr(\lambda)&=&\frac{\pi}{12}\frac{\rho_t^0(q)}{[\ezp]^2}
\left\{ {\left[\frac{u(q)}{2\pi}-\frac{\ezpp}{\ezp}
+\frac{{\rho_t^0}^\prime(q)}{{\rho_t^0}(q)}\right] u(\lambda)
-w(\lambda)} \right\}+{\cal O}(T)\ ,
\label{diffr2}
\eea
with
\be
w(\lambda)=K^\prime(\lambda,q)+K^\prime(-\lambda,q)+\frac{1}{2\pi}
\int_{-q}^{q}d\mu\ K(\lambda,\mu)\ w(\mu)\ ,
\ee
where $K^\prime(\lambda,q)=dK(\lambda,q)/d\lambda$.
Finally, the low-temperature expansion for $v(\lambda)$ is found to be
\be
v(\lambda)=\frac{{\eps^0}^\prime(\lambda)}{2\pi\rho_t^0(\lambda)}
-\frac{1}{24\ezp}\ \frac{u^\prime(\lambda)}{\rho^0_t(\lambda)}\
T^2-\frac{{\eps^0}^\prime(\lambda)
\dr(\lambda)}{2\pi[\rho_t^0(\lambda)]^2} T^2
+{\cal O}(T^3)\ .
\label{vT}
\ee

\subsection{Asymptotics of the correlator at small temperatures}

Having obtained the leading corrections (in temperature) of
$\eps(\lambda)$, $\rho_t(\lambda)$ and $v(\lambda)$ we are now in a
position to determine the leading temperature correction to the
conformal result \r{asym2} for general values of the coupling $c$.
In order to do so we need to expand the integral in (\ref{asym3}),
which is of the form
\be
\chi=\frac{1}{2\pi}\int_{-\infty}^{\infty}d\lambda\ f(\lambda)
\ln\left|\frac{e^{\frac{\eps(\lambda)}{T}}-1}
{e^{\frac{\eps(\lambda)}{T}}+1}\right|\ ,
\ee
where $f(\lambda) = \frac{|x-v(\lambda)t|}{2\pi\rho_t(\lambda)}$. 
The leading contributions to $I(T)$ come from the vicinity of the
points $\pm q_T$, where $\eps(\lambda)$ vanishes. Expanding
$f(\lambda)$ in powers of $T$, using \r{eT}, \r{diffr2} and \r{vT},
and integrating around $\pm q_T$ we find
\be
\chi= A T + B T^3 +{\cal O}(T^4)\ .
\ee
We find that $A$ reproduces the ``conformal'' result \r{asym2}. There
is no correction to order $T^2$. $B$ is a complicated expression as it
reflects the operator content of the theory so that we give an
explicit expression only in the case $x\gg v_F t$. 

\bea
B&=&
-\frac{\pi^3Z^\prime\ezpp}{8Z^2(\ezp)^4}
-\frac{\pi^3{Z^\prime}^2}{24Z^3(\ezp)^3}
+\frac{\pi^3Z^{\prime\prime}}{48Z^2(\ezp)^3}
-\frac{\pi^3(\ezpp)^2}{8Z(\ezp)^5}\nn
&+&\frac{\pi^2u(q)\ezpp}{24Z(\ezp)^4}
-\frac{\pi^2u^\prime(q)}{24Z(\ezp)^3}
+\frac{\pi^2u(q)Z^\prime}{24Z^2(\ezp)^3}
+\frac{\pi^3\ezppp}{24Z\ezp}
+\frac{\pi^2\Delta\rho(q)}{Z^2\ezp}\ ,
\eea
where $Z^\prime=2\pi{\rho_t^0}^\prime(q)$ and 
$Z^{\prime\prime}=2\pi{\rho_t^0}^{\prime\prime}(q)$.

\begin{figure}[ht]
\begin{center}
\noindent
\epsfxsize=0.65\textwidth
\epsfbox{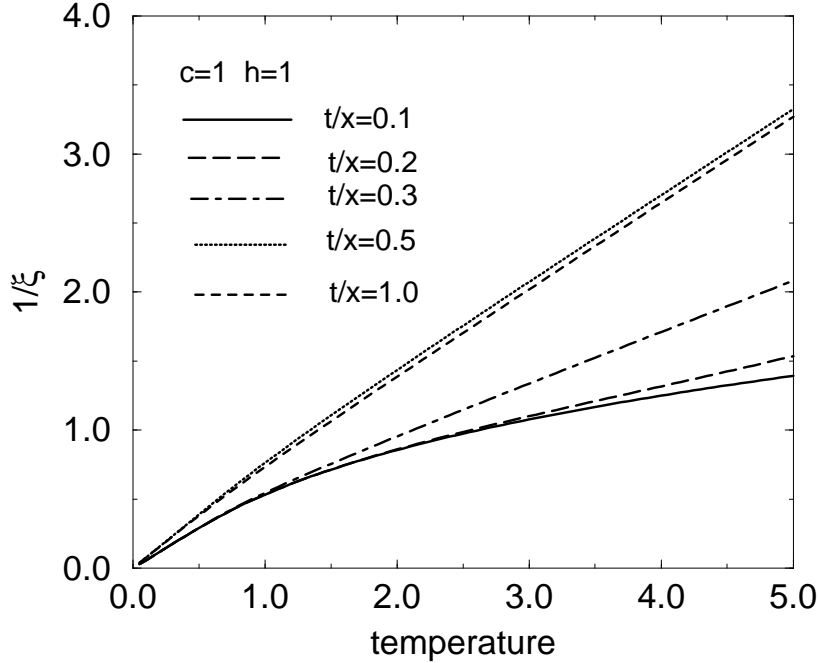}
\end{center}
\caption{\label{fig:c1}%
Inverse correlation length as a function of temperature at $c=1$,
$h=1$.  
} 
\end{figure}

At higher temperatures we again resort to a numerical solution of the
relevant integral equations and integrals. In Fig. \ref{fig:c1} we
plot the inverse correlation length \r{length} as a function of
temperature for $c=1$ and $h=1$. 

We obtain a qualitatively similar picture to the impenetrable
case. Once again the exponential decay of correlations is fastest in
the direction $x=v_Ft$. However, the dependence on $t/x$ is more
pronounced than in the impenetrable case. This particularly clear in
the space-like regime.

\begin{figure}[ht]
\begin{center}
\noindent
\epsfxsize=0.65\textwidth
\epsfbox{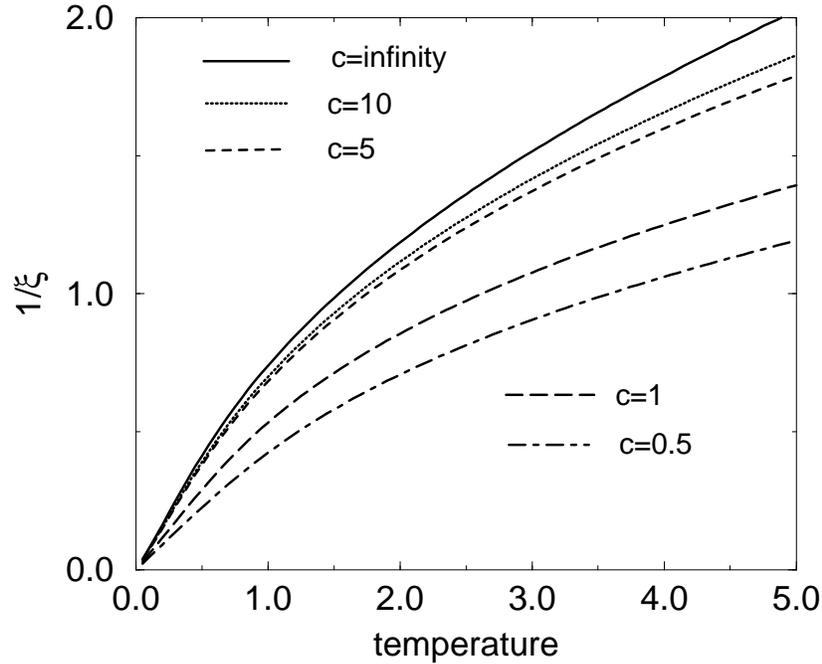}
\end{center}
\caption{\label{fig:plotc}%
Inverse correlation length for several values of $c$ as a function of
temperature at $h=1$.  
} 
\end{figure}

\section{Conclusions}
We have studied the rate of exponential decay of finite-temperature
dynamical correlation functions of local fields in the
$\delta$-function Bose gas. We have explicitly evaluated corrections
to conformal behaviour at low temperatures.
For impenetrable bosons the results are qualitatively the same, and in
addition we are able to get all the terms in an asymptotic series
for the low-temperature behaviour of the rate of decay.

\end{document}